\pdfoutput=1
\documentclass[11pt,a4paper,superscriptaffiliations]{article}
\usepackage{graphicx}
\usepackage{siunitx}
\usepackage{cite}
\usepackage{authblk}
\usepackage[margin=2.8cm]{geometry}

\begin{document}

\title{Non-planar femtosecond enhancement cavity for VUV frequency comb applications}

\author[1,2]{Georg Winkler}
\author[1,2]{Jakob Fellinger}
\author[1,2]{Jozsef Seres}
\author[1,2]{Enikoe Seres}
\author[1,2]{Thorsten Schumm}
\affil[1]{Vienna Center for Quantum Science and Technology}
\affil[2]{Atominstitut, TU Wien, Stadionallee 2, 1020 Vienna, Austria}
\renewcommand\Affilfont{\itshape\small}

\date{}

\maketitle

\begin{abstract}
External passive femtosecond enhancement cavities (fsECs) are widely used to increase the efficiency of non-linear conversion processes like high harmonic generation (HHG) at high repetition rates. Their performance is often limited by beam ellipticity, caused by oblique incidence on spherical focusing mirrors. We introduce a novel three-dimensionally folded variant of the typical planar bow-tie resonator geometry that guarantees circular beam profiles, maintains linear polarization, and allows for a significantly tighter focus as well as a larger beam cross-section on the cavity mirrors. The scheme is applied to improve focusing in a Ti:Sapphire based VUV frequency comb system, targeting the 5th harmonic around 160\,nm (7.8\,eV) towards high-precision spectroscopy of the  low-energy isomer state of Thorium-229. It will also be beneficial in fsEC-applications with even higher seeding and intracavity power where the damage threshold of the mirrors becomes a major concern.
\end{abstract}


\section{Introduction}
Since their invention at the turn of the century, optical frequency combs have quickly proven a revolutionary tool for precision measurements of optical frequencies, paving the way for next-generation technologies like optical clocks~\cite{evolving_comb}. 
Currently, femtosecond laser sources which are the basic building block of these universal tabletop devices, are only available directly in visible to near-infrared spectral regions owing to the limited choice of suitable gain media. Therefore extending the accessible wavelength range via non-linear conversion processes is highly desirable and an active field of research.

One particularly exciting application in this respect could be the realization of an optical clock based on the radioisotope thorium-229 which is expected to possess an optically accessible nuclear transition at around 160\,nm in the vacuum ultraviolet (VUV)~\cite{beck2007}. The combination of high frequency, long lifetime, and other advantageous properties linked to the nuclear nature of the transition, suggest it as a candidate for a clock of unprecedented accuracy~\cite{peik2008, kuzmich, kazakov2012, thcrystal}. 

The method of choice to attain these and shorter wavelengths is high-harmonic-generation (HHG) in gaseous media which requires a high peak power density of the fundamental radiation well above $\SI{e13}{W/cm^2}$. However, in this regime usual amplification techniques like chirped pulse amplifiers (CPA) or regenerative amplifiers typically imply a significant reduction of the pulse repetition rate (originally in the range of $\SI{100}{MHz}$). This renders the comb spectrum unusable for the intended frequency measurements as it reduces the spectral separation of the individual comb lines. 

One obvious measure to increase the power density in the non-linear medium is tight focusing. The second measure is to recycle the unconverted pulse fraction in an external femtosecond enhancement cavity (fsEC). This applies to second harmonic generation (SHG)~\cite{shg}, optical parametric generation and amplification (OPG, OPA) extending combs into the mid-infrared region~\cite{opa0, opa1, opa2}, as well as HHG which has enabled frequency combs reaching far into the extreme ultraviolet (XUV)~\cite{vancouver_review, kobayashi, arizona, jones, gohle, seres2012}.

Due to the passive nature of such cavities and the necessity for precise dispersion management of the broadband pulses, only reflective optics can be used and as little material as possible should be introduced into the cavity. Astigmatic aberrations, introduced by oblique incidence on the focusing mirrors, can not be readily compensated by Brewster plates. This typically leads to elliptical beam profiles and restricts both the minimum attainable focus size as well as the maximal beam cross-section on the cavity mirrors.

Minimizing the beam waist in order to reach the necessary peak power density is especially important when low seeding power is available. 
This applies to the above-mentioned application of a Th-229 clock, which requires a Ti:Sa oscillator as a seed laser, offering an average power orders of magnitude below what is available with fiber-based combs. On the other hand, for such high-power systems, increasing the beam diameter on the resonator optics becomes the main concern in order to stay below the damage thresholds of the dielectric coatings~\cite{largemode}. The achievable maximum value is reduced in particular by the fact that larger beam diameters necessitate larger incidence angles on the focusing mirrors, increasing the amount of unavoidable astigmatism. Several options have been proposed to address these problems, including additional cylindrical mirrors~\cite{largemode}, parabolic focusing mirrors~\cite{parabolic}, or non-planar cavities~\cite{nonplanar}.

In this study we introduce and characterize a straightforward non-planar configuration of the standard bow-tie cavity configuration, which significantly relaxes the before-mentioned aberration-related limitations, without introducing significant additional complexity or additional elements. The design is successfully used to generate a frequency comb at \SI{160}{nm} via HHG, within the framework of the thorium-229 nuclear clock project.

\section{Resonator design}
\begin{figure}[htpb]
\centering\includegraphics[width=13.28cm]{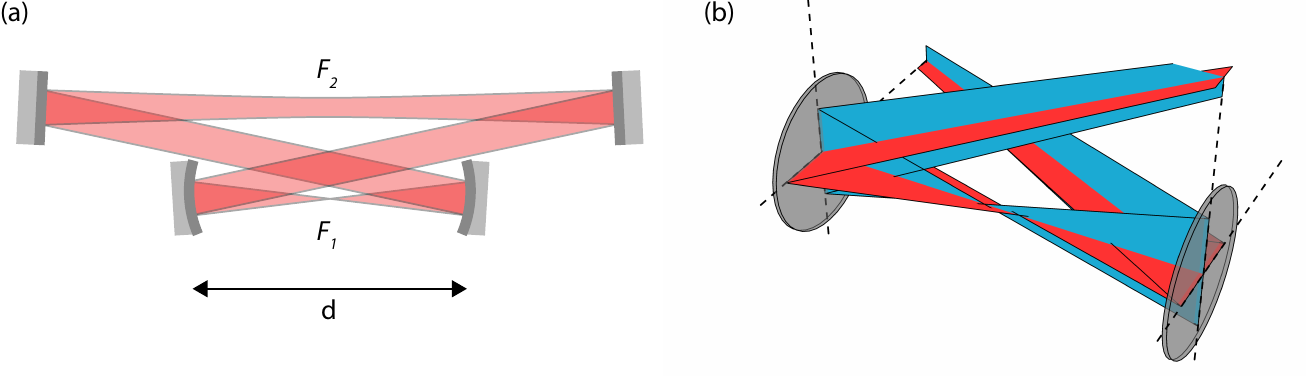}
\caption{Comparison of planar and three-dimensional resonator geometry. (a) Bow-tie resonator in standard planar layout. The primary and secondary foci are labeled $F_1$ and $F_2$. (b)~Detailed view of the orthogonally folded beams around the two focusing mirrors of the non-planar geometry, with vertical (blue) and horizontal (red) optical planes.}
\label{fig:cavity}
\end{figure}

The typical fsEC for high repetition rate, high-power non-linear conversion applications like HHG is the bow-tie cavity similar to Fig.~\ref{fig:cavity}(a), consisting of two identical spherical mirrors and at least two folding mirrors in a symmetric planar arrangement.
As a stable resonator it fulfils the requirement for significant power enhancement. 
Implementation as a ring-resonator avoids a double pass in opposite directions through the non-linear material, eliminates optical feedback into the source and allows straightforward separation of the reflected beam from the seeding beam to derive a locking signal.
The bow-tie type folding of the intracavity beam path allows small incidence angles on the focusing mirrors which, together with the symmetry of the setup, helps to minimize astigmatic and other optical aberration effects. 

\begin{figure}[htpb]
\centering\includegraphics[width=13.28cm]{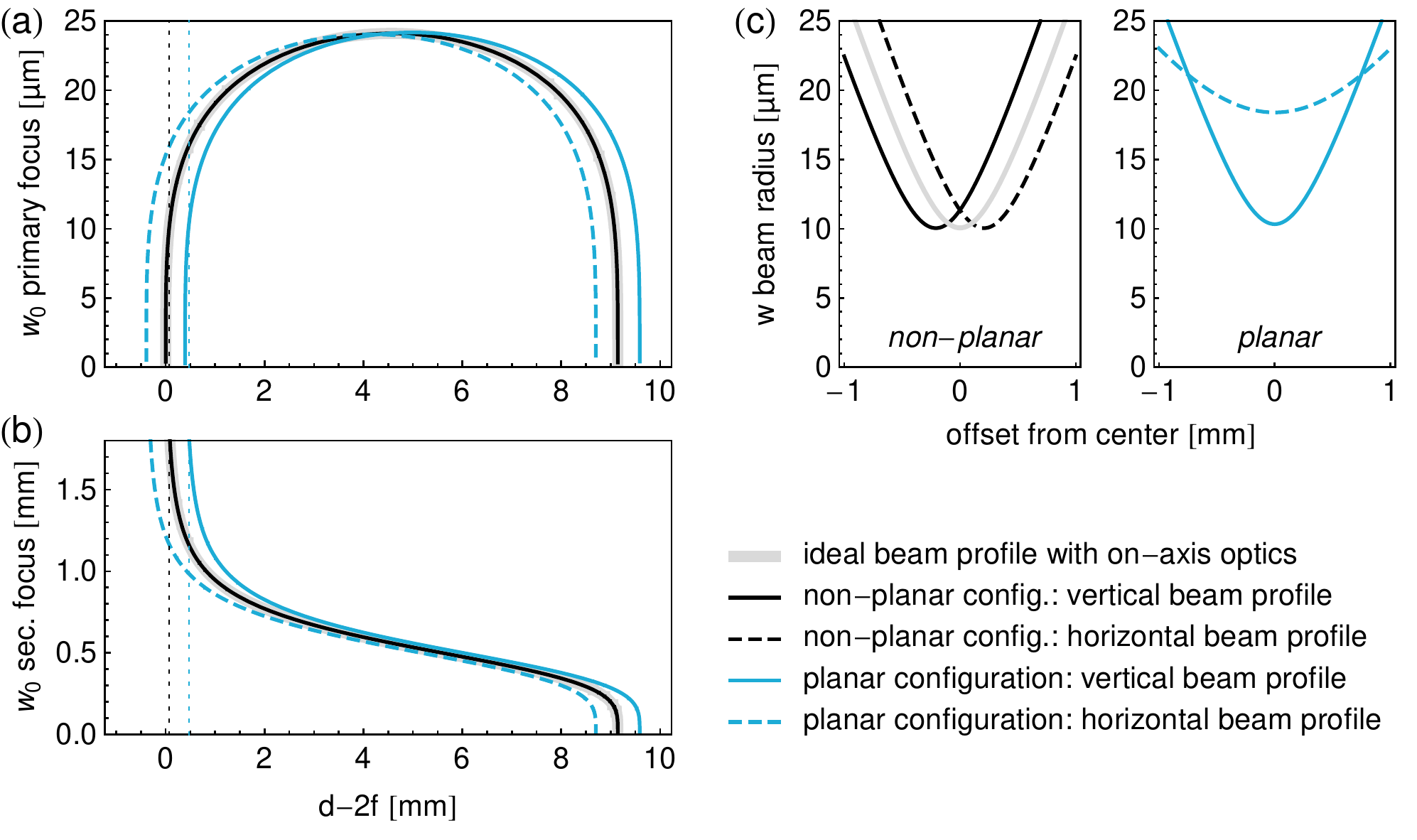}
\caption{Stability diagram and focus profiles. The calculation is based on the experimental parameters given in section \ref{sec:setup}, that is $f=\SI{75}{mm}$, $\lambda = \SI{800}{nm}$, $L=\SI{2.77}{m}$, $AOI=\SI{4.2}{\degree}$. (a)~The primary and (b)~secondary beam waist size is plotted over the entire stability region. The values for the horizontal and vertical plane differ in case of the planar layout (blue), whereas they are the same for the non-planar one (black), coinciding with the value for the ideal case of zero incidence angle (grey). (c)~The corresponding Gaussian beam profiles are plotted at comparable positions close to the stability edge, indicated by dotted lines.}
\label{fig:stability3D}
\end{figure}

Stable Gaussian modes can be derived using ray transfer analysis, by requiring the corresponding complex beam parameters $q$ to reproduce themselves after multiplication by the ABCD matrix corresponding to one cavity roundtrip~\cite{kogelnik}. 
The generic solutions
\begin{equation}
\frac{1}{q} = \frac{D-A}{2B}-i\frac{\sqrt{4-(A+D)^2}}{2|B|}
\end{equation}
are physically meaningful (leading to real-valued beam radii) if the square root is real. This stability criterion is usually stated as  $-1<\frac{1}{2}(A+D)<1$ and follows directly from comparison with the relations linking $q$ to the beam radius $w$, wavefront radius of curvature $R$, waist radius $w_0$ and waist position $z_0$:
\begin{equation}
\frac{1}{q}=\frac{1}{R}-i \frac{\lambda}{\pi w^2}
\end{equation}
\begin{equation}
q=(z-z_0)+i \frac{\pi w_0^2}{\lambda}.
\end{equation}
For the bow-tie cavity this implies a single interval of allowed values for the separation $d$ of the focusing mirrors, defining the stability region, for a given cavity length $L$ (determined by the seeding laser repetition rate) and a choice of focal length $f$ for the two identical spherical focusing mirrors. The dependence of the two intracavity beam waists on $d$ is plotted in Figure~\ref{fig:stability3D}. The general behaviour is best considered for the idealized theoretical case of orthogonal incidence on the focusing mirrors (gray curve), which would be equivalent to the use of on-axis refractive optics: The spot size of the primary focus ($F_1$ in Fig.~\ref{fig:cavity}(a)) asymptotically approaches zero at both edges of the stability range, only limited by the finite size of the cavity mirrors. In the secondary cavity arm the beam waist shrinks as well on the outer stability edge (where $d$ is larger), forming a pronounced but weaker secondary focus ($F_2$), but diverges on the inner edge, leading to a more or less collimated beam. 
It should also be noted that the cavity alignment sensitivity has been shown to diverge on the outer edge~\cite{largemode}.

For the real-world scenario of finite incidence angles, Gaussian beam propagation has to be considered separately for the horizontal and vertical plane, since the effective focal lengths assume different values of $f_{\parallel}=f \cdot \cos \theta$ and $f_{\perp}=f / \cos \theta$ respectively with $\theta$ denoting the angle of incidence (AOI). This leads to shifted stability limits for the two planes (blue curves), narrowing the overall accessible stability range and causing increasingly elliptical beam profiles as the stability edges are approached. 
As can be seen from the plots of the beam profiles in Figure~\ref{fig:stability3D}(c), the astigmatic aberration manifests itself as pure ellipticity since the positions of the beam waists in both planes still overlap, which obviously follows from the symmetry of the mirror arrangement.

The basic idea put forward in this work is to counteract these limitations by rotating the plane of incidence for one of the focusing mirrors to the vertical plane by 90 degrees (Fig.~\ref{fig:cavity}(b)), so that for both horizontal and vertical mode equations there is one mirror with increased and one with reduced effective focal length to consider~\cite{yefet1013}. 
This way the beam waists in both planes are of the same size, however for both planes the symmetry of the setup around the midpoint between the two focusing mirrors is broken, slightly shifting the position of the beam waists along the optical axis - in different directions for different planes, due to switched order of appearance of the effective focal lengths. This again introduces a small astigmatism. 

\begin{figure}[htpb]
\centering\includegraphics[width=13.28cm]{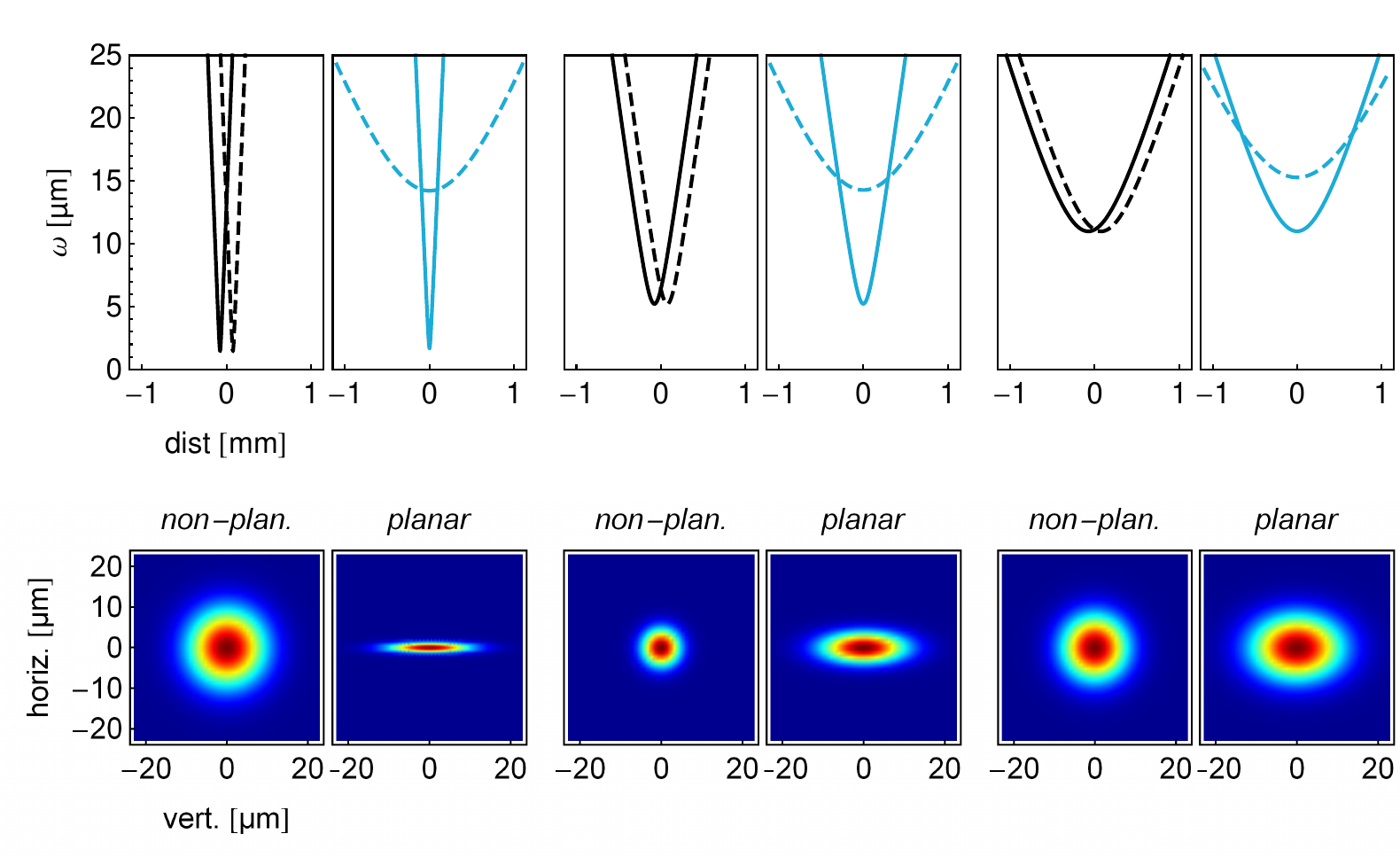}
\caption{Comparison of primary focus geometries for increasing distance ($<1\,\mu m$, $5\,\mu m$ and $0.1\,mm$ respectively) from the stability edge. The beam profiles are plotted solid for the vertical plane and dashed for the horizontal plane. The bottom cross sections apply to the symmetry point at position zero. All calculations are done based on the same parameters as in Fig.~\ref{fig:stability3D}, except for a more typical AOI of \SI{2.5}{\degree}. The middle pair represents a configuration where the non-planar setup is performing optimally.}
\label{fig:optimum3D}
\end{figure}

The crucial advantage is that now the beam profiles are always circular both at the midpoint where the circle of least confusion occurs, as well as - to very good approximation - further away from the focus. Compared to the elliptical beam profiles of an equivalent planar setup, the power density is decreased on the cavity mirrors and increased in the focus under typical conditions (Fig.~\ref{fig:stability3D}). The ratio of the two cross-sections, signifying the amount of focusing improvement, reaches a maximum at a position very close to the stability edge. Only when moving even closer to the stability edge (in which case the cavity would be hard to stabilize anyway), will the astigmatism become noticeable. Since the separation distance of the beam waists in the different planes remains almost constant over the entire stability range, degradation of the focusing geometry is mainly due to the larger divergence of an even tighter focus (Fig~\ref{fig:optimum3D}). The separation distance shrinks for smaller angles of incidence and shorter focal lengths, so that minimizing those can still be beneficial to attain the smallest possible focal spot. In general this does not mean that the non-planar scheme becomes less effective when moving to longer focal lengths, since also the beam ellipticity exhibited by the equivalent planar setup would increase in this case, at comparable points in the stability diagram.

For the enhancement cavity to be practically useful, the (linear) polarization state of the resonating field has to be conserved, which is not guaranteed for arbitrary non-planar layouts, since the polarization main axes before and after a reflection are defined with respect to the plane of incidence, which can easily lead to geometric rotation of the polarization after a single roundtrip. As detailed in the following section, the resonator presented here has been folded in a way to preserve the polarization similar to a planar one.

\section{Experimental Setup}
\label{sec:setup}
We seed our non-planar fsEC with a Ti:Sapphire oscillator-based frequency comb, generating $\SI{25}{fs}$-long pulses centred around \SI{800}{nm} with \SI{900}{mW} of average power at a repetition rate of \SI{108}{MHz} (FC8004, Menlo Systems). This system was chosen in order to directly obtain the \SI{160}{nm} target wavelength as the fifth harmonic of the fundamental wavelength.

\begin{figure}[htpb]
\centering\includegraphics[width=13.28cm]{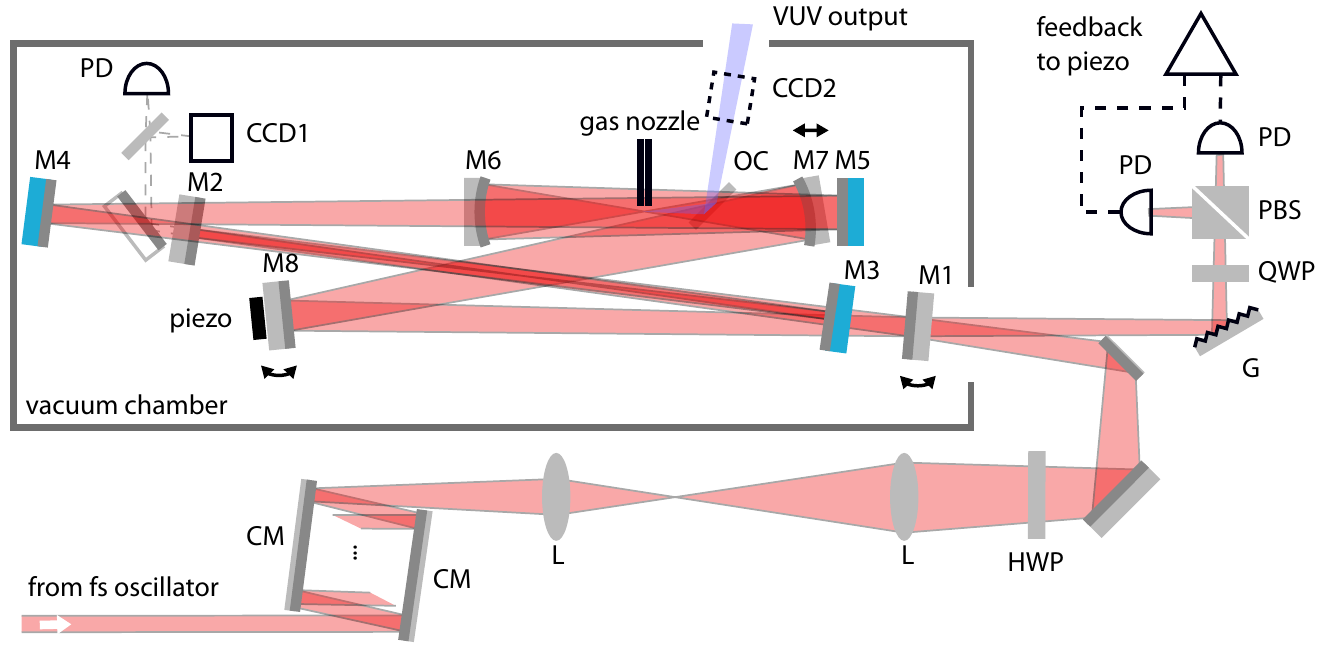}
\caption{Schematic of the fsEC experimental setup. M1-M8: cavity mirrors (those depicted in blue are aligned on a second horizontal level), CM: chirped mirrors, L: lenses for mode-matching, PD: photodiode, OC: output coupler, PBS: polarizing beam splitter, QWP: quarter wave plate, HWP: half wave plate, G: grating.}
\label{fig:setup}
\end{figure}

The cavity length of \SI{2.77}{m} is matched to the seeding laser repetition rate and brought to a manageable footprint using 6 folding mirrors (M1-M5, M8 in Fig.~\ref{fig:setup}). 
To fulfil the requirement of polarization conservation in a straightforward way, all mirrors are arranged in two horizontal planes and whenever a segment of the beam path is connecting mirrors at different height, it is made sure that the preceding and subsequent segments lie in the same vertical plane. Therefore a set of four mirrors aligned in a vertical plane is required to raise (M1-M4) or lower (M4-M7) the beam in a Z-shape. 

The group delay dispersion (GDD) of all mirrors is chosen to yield zero GDD per roundtrip. This was verified by checking the symmetry of the resonance peaks and by observing the full seed spectrum being coupled into the cavity (Fig.~\ref{fig:hhg}(a)). All mirrors are broadband high reflectors, except for the input coupling mirror M1 with a transmission of one percent.

A two-lens telescope is used for mode matching to the fsEC, focusing the input beam to match the secondary intracavity waist at its correct position between mirrors M2 and M3 (focal lengths of the lenses: \SI{1000}{mm},  \SI{750}{mm}; distance of the first lens to the femtosecond oscillator: \SI{920}{mm}; distance between the lenses: \SI{2430}{mm}; distance of the second lens to the secondary intracavity waist: \SI{890}{mm}). The secondary focus can be easily monitored by picking up the leak signal of mirror M2. Since the focus is automatically reproduced outside the cavity this way, it can be directly imaged on a CCD sensor (CCD1) without the need for further imaging optics which might introduce aberrations. The scheme proved especially helpful for very precise geometric alignment of the cavity mirrors. 

To compensate the cumulative GDD of about \SI{570}{fs^2} added by mode-matching optics, vacuum chamber window, incoupling mirror substrate and propagation in air, the Fourier-limited pulses are pre-compressed via several reflections on chirped mirror pairs (see Fig.~\ref{fig:setup}).

The cavity is built inside a vacuum chamber to avoid reabsorption of the generated VUV signal by oxygen in air. All mechanical mounts requiring adjustment during operation are remote controlled via direct drive piezoelectric steppers. The driver electronics were custom-built to minimize single-step overshoot for most precise alignment.
Acoustic vibrations picked up by the floated optical table in the noisy lab environment proved to be the dominant contribution to lock instability, as compared to residual vibrations from the vacuum pumps. For this reason all cavity components are mounted on a separate breadboard resting on sorbothane rubber feet for additional vibration damping inside the vacuum chamber, which itself is directly mounted to the optical table.

A small nozzle with \SI{100}{\micro m} inner diameter, mounted on a 3D translation stage, is used to direct a Xenon gas jet at the strong primary intracavity focus between the two concave mirrors (M6 and M7) with \SI{75}{mm} focal length. 

A \SI{100}{\micro m} thick BK7 glass plate is placed after the focus at Brewster angle (for the fundamental wavelength) to couple out generated harmonics. 
Spectrometers for the visible and VUV (McPherson Model 234, with detector choice of Andor CCD camera DO940P-BN-995 or Hamamatsu photomultiplier R6836) as well as a beam profiler are connected to the output port for diagnostics. 
To be able to experiment with other more bulky outcoupling elements, a rather large angle of incidence of \SI{4.2}{\degree} on the focusing mirrors is chosen. 

One cavity mirror (M8) is mounted on a piezo to lock the cavity length to the repetition rate of the Ti:Sa oscillator, using a H\"ansch-Couillaud scheme~\cite{haenschlock} to derive the error signal, exploiting the presence of the outcoupling plate as a strongly polarization selective element. The carrier envelope offset of the oscillator is stable enough to be manually adjusted for optimal power enhancement.

\section{Results}
\begin{figure}[]
\centering\includegraphics[width=13.28cm]{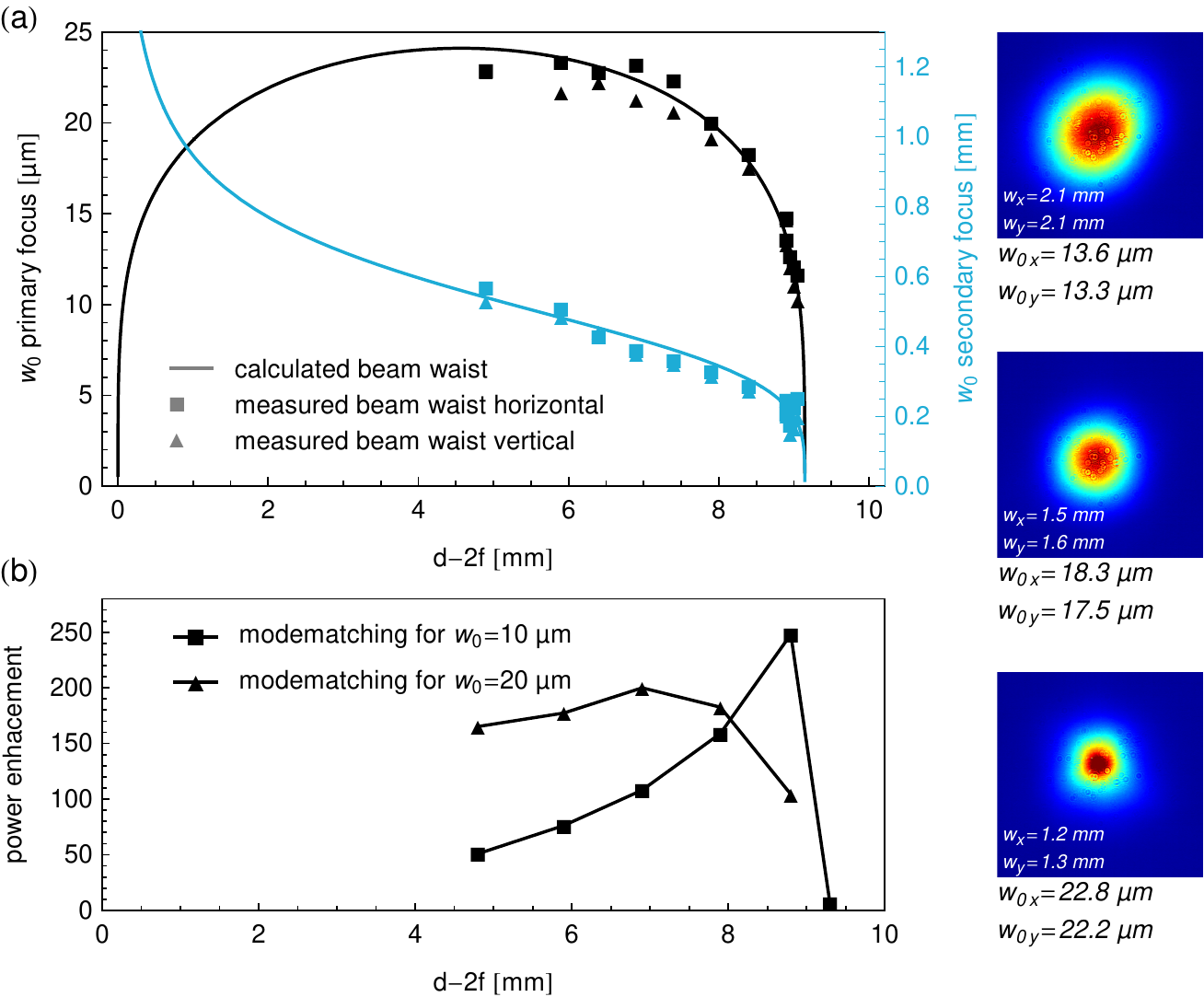}
\caption{Measured beam waists in vertical and horizontal planes for (a)~both intracavity foci and (b)~corresponding power enhancement. The enhancement factors are given for two different input beam configurations which were kept constant during the measurement. The depicted beam cross sections were measured at a constant distance of \SI{11}{cm} from the primary focus.}
\label{fig:beamwaists}
\end{figure}

We locked the frequency comb to the enhancement cavity at different positions in the stability zone (by varying the focusing mirror separation distance $d$) and measured the beam waist geometry of both intracavity foci as well as the intracavity power. The secondary focus was directly imaged onto CCD1 via the convergent beam leaking from mirror M2  before the focus. The primary focus was indirectly determined by recording a profile of the divergent beam at known distance after the focus on CCD2 
and calculating the corresponding Gaussian beam waist. The measured beam waists (Fig.~\ref{fig:beamwaists}(a)) show good agreement with the theoretical prediction. We were able to produce a focus with a $1/e^2$ radius as small as $\SI{10}{\micro m}$ and an average ellipticity $1-w_x/w_y$ less than 6 percent, although using a rather large AOI of $\SI{4.2}{\degree}$ on the focusing mirrors. 

An enhancement factor of 250 was reached  (Fig.~\ref{fig:beamwaists}(b)), defined as the ratio of circulating intracavity power $P_{circ}$ to incoming power $P_{in}$. It is related to the incoupling mirror transmission $T$ and the round trip power loss factor $L'$ accounting for all other losses (absorption, scattering, diffraction and leakage of all mirrors except the incoupling mirror) by
\begin{equation}
\frac{P_{circ}}{P_{in}} = \eta_{spectral} \cdot \eta_{spatial} \cdot \frac{ T}{(1-\sqrt{(1-T)(1-L')})^2}.
\label{eq:enh}
\end{equation}
The coupling efficiencies $\eta$ quantify the amount of spectral and spatial overlap of the comb and cavity modes.
As follows from Eq.~(\ref{eq:enh}), the measured enhancement is already close to the theoretical limit of about 400, given an input coupler transmission $T$ of 1 percent. The pronounced peak in the measured enhancement curves is due to the fact that (spatial) mode matching was not reoptimized for every datapoint. 

\begin{figure}[htpb]
\centering\includegraphics[width=13.28cm]{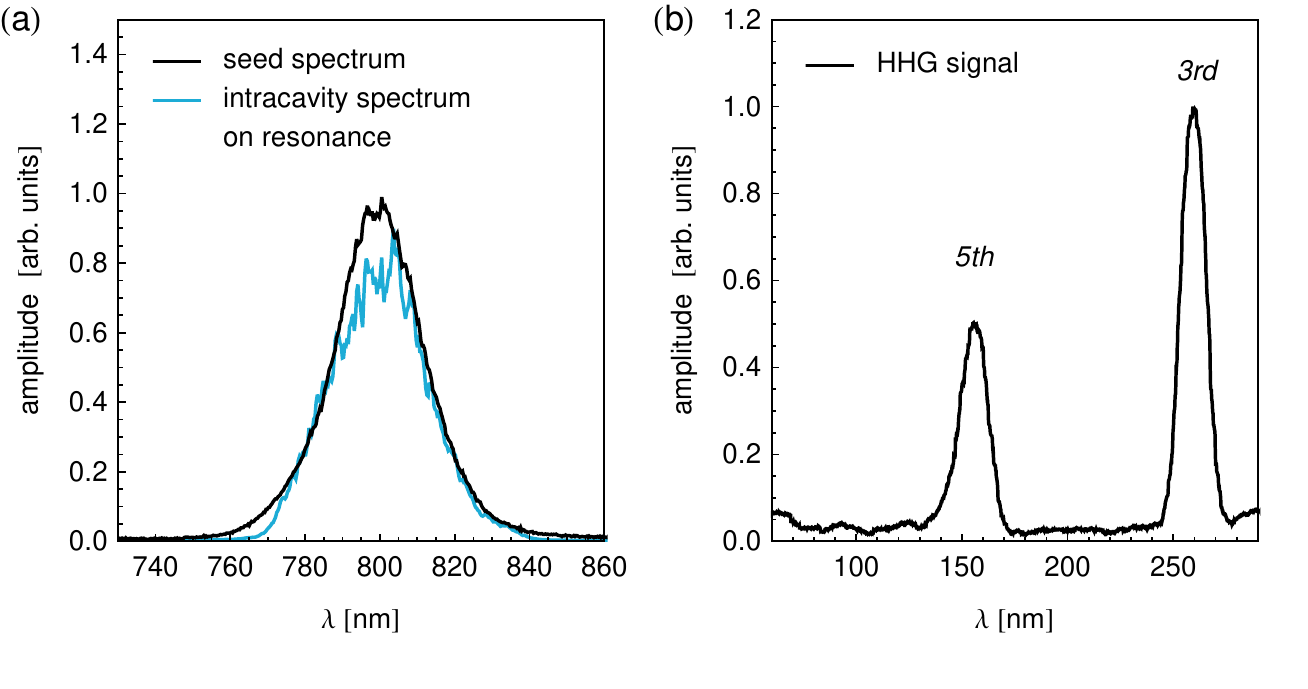}
\caption{Spectrum of the (a)~fundamental mode and (b)~harmonic output. Amplitudes of outcoupled harmonics are not to scale for the latter, due to different filtering and detection efficiencies of the spectrometer.}
\label{fig:hhg}
\end{figure}

For best spectral overlap, the cavity length was actively locked to the comb repetition rate, while the carrier-envelope offset (CEO) frequency of the comb was manually tuned for maximum intracavity power. The fact that almost the full input spectrum could be coupled into the cavity (Fig.~\ref{fig:hhg}(a)) shows that equidistant cavity mode spacing could be achieved by selecting mirror coatings with zero net round trip GDD. Deviation from this condition (caused by added GDD due to changing background pressure for example) could also be observed conveniently through increasing asymmetry of the cavity resonance peaks when scanning the cavity length. 

The main contribution to intracavity loss came from residual reflection and scattering at the outcoupler plate. 
Considering the measured maximum enhancement factor and assuming $\eta_{spectral}=\eta_{spatial}\approx1$ it follows from Eq.~(\ref{eq:enh}) that $L'\approx 0.3 \%$, suggesting a potentially improved enhancement by impedance matching the cavity with an incoupling mirror transmission of equal magnitude. Nevertheless it was decided to deliberately accept a slightly decreased enhancement in the overcoupled regime ($T>L'$) in order not to risk entering an undercoupled regime ($T<L'$) with much stronger associated penalty on intracavity power, when $L'$ increases during operation as a consequence of the nonlinear effect and ionization in the gas used for HHG.

With the xenon jet at the primary focus switched on, a fifth harmonic signal at \SI{160}{nm} was successfully generated (Fig.~\ref{fig:hhg}(b)). The yield was optimized through variation of the xenon backing pressure, alignment of the nozzle position in three translational degrees of freedom and adjustment of the CEO frequency of the comb. 

\section{Conclusions}
We have developed a scheme to compensate beam ellipticity in femtosecond enhancement cavities in standard bow-tie configuration, by moving to a non-planar layout that still conserves the polarization throughout the entire cavity. Our solution introduces manageable additional complexity with respect to setup and operation, and can be built from the same readily available off-the-shelf components used in the planar setup. The freedom to tune the intracavity focus spot size over a wider range and to reduce the pulse energy density on the cavity mirrors, is beneficial in non-linear conversion applications with both low- and high-power seed laser source. 

Using the new design, we realized a Ti:Sapphire laser seeded femtosecond enhancement cavity with about 30\,fs intracavity pulse duration and overall enhancement factor of 250. We demonstrated tunability of the beam waist in the intracavity focus between \SI{10}{\micro m} and \SI{22}{\micro m} while preserving the high enhancement factor of the cavity. In a proof-of-principle experiment, the fifth harmonic at 160\,nm was generated in the intracavity focus in Xe gas. Realization of such a VUV source is an essential step e.g. for the precise measurement of optical transition of the \textsuperscript{229}Th nucleus or for the realization of an \textsuperscript{115}In+ optical clock \cite{wakui}.

\section*{Acknowledgments}
This work was supported by the FWF Project Y481 and the ERC project 258604-NAC. GW acknowledges discussion with J. Weitenberg and A. Ozawa on large-mode cavities and issues of alignment sensitivity.

\end{document}